\documentclass[pra,bibnotes,twocolumn]{revtex4}
\usepackage{graphicx}
\begin{document}
\draft
\def\ds{\displaystyle}
\title{Edge states at the interface of non-Hermitian systems}
\author{C. Yuce}
\address{Department of Physics, Anadolu University, Turkey }
\email{cyuce@anadolu.edu.tr}
\date{\today}
\begin{abstract}
Topological edge states appear at the interface of topologically distinct two Hermitian insulators. We study the extension of this idea to non-Hermitian systems. We consider PT symmetric and topologically distinct non-Hermitian insulators with real spectra and study topological edge states at the interface of them. We show that PT symmetry is spontaneously broken at the interface during the topological phase transition. Therefore topological edge states with complex energy eigenvalues appear at the interface. We apply our idea to a complex extension of the Su-Schrieffer-Heeger (SSH) model.
\end{abstract}
\maketitle

\section{Introduction}

 A topological insulator has gapless edge states and gapped spectrum in the bulk. The most interesting feature of topological edge states is its robustness against symmetry protected disorder. Topological phase is a very general concept that can be applied to many branches of physics. For example, topological phase in photonics has attracted a great deal of attention in the last decade. In recent years, topological photonics systems with gain and loss becomes a rapidly growing field of study. This is particularly interesting since theories of the non-Hermitian extension of topological phase can be tested in photonics. Quantum mechanical systems are generally described by a Hermitian Hamiltonian while gain can be implemented and loss is generally inevitable in optics. Topological photonics can have also some interesting technological applications such as topological lasing \cite{lasing}. A topological laser is a laser that is immune to disorder and fabrication defects. Topological phase in Hermitian systems have been studied extensively and the periodical table of topological insulators for Hermitian Hamiltonians is well known. However, there had been hot debate about the existence of topological phase in non-Hermitian systems until recently \cite{PTop2,PTop3,ekl56,PTop4,hensch,PTop1}. We note that Berry phase can not be directly generalized to non-Hermitian systems \cite{sbt5}. Some authors came with an idea that topological phase is not compatible in the $\mathcal{PT}$ symmetric region, where $\mathcal{P}$ and $\mathcal{T}$ are parity and time reversal operators, respectively. Some other authors found either growing or decaying edge states in topological domain. Until 2015, there had been few paper on this topic because it was generally believed that topological phase is absent in non-Hermitian systems. The first paper that theoretically predicted stable topological phase in a non-Hermitian Aubry-Andre model appeared in the literature in 2015 \cite{cem0001}. In a year, an experiment was realized \cite{sondeney1} and topological edge states in lossy waveguides were observed through fluorescence microscopy. The first experimental realization of topological edge states is followed by many other papers on non-Hermitian topological phase and this subfield becomes a rapidly growing field. So far, most of investigations in this field have been restricted to one dimensional problems \cite{sbt1,sbt3,sbt6,sbt11,sbt8}. Of special importance is the paper \cite{sbt2} which states that topological insulating phase can also be realized only by gain and loss. Theories on $1D$ topological phase was generally constructed on a generalized non-Hermitian Su-Schrieffer-Heeger (SSH) model \cite{sbt4,cmyc}. On contrary to $2D$ topological insulators whose edge modes are propagating either chiral or helical modes depending on the topological invariant, edge modes in $1D$ are accumulated at edges and decays rapidly away from edges. In the literature, there are some papers studying non-Hermitian topological insulator in $2D$, too \cite{sbt100}. Not only topological insulators but also topological superconductors and Majarona modes have been studied in non-Hermitian systems \cite{sbt12,sbt16,sbt17,sbt7,sbt18}. It is well known that standard classification of topological insulators and superconductors according to the three discrete symmetries for a given Hamiltonian in any dimension fails if the system is time-dependent. Another kind of topological insulators that appear in time-periodic systems are called Floquet topological insulators \cite{flotop1,flotop11}. Non-Hermitian Floquet topological phase was also studied in \cite{sbt10,sbt15}.  \\
This subfield is new and there are some open problems in non-Hermitian topological systems such as bulk-boundary correspondence in non-Hermitian systems \cite{brt0,brt1} and non-Hermitian topological invariants. Standard bulk-boundary correspondence tells us about symmetry protected edge states at the interface of two topologically distinct Hermitian systems. At the interface of two topologically inequivalent systems, there exists gapless conducting edge states although the two systems are insulators. The bulk energy gap closes somewhere along the way when the topologically nontrivial system is in contact with a topologically trivial one. A question arises. What are the properties of topological edge states at the interface of topologically distinct two non-Hermitians systems? Are they robust against disorder? Is $\mathcal{PT}$ symmetry spontaneously broken during topological phase transition in non-Hermitian systems? In this paper, we will discuss what happens if topologically distinct two non-Hermitian Hamiltonians have an interface. We will discuss that topological edge states with complex energy appear at the interface of topologically distinct two non-Hermitian Hamiltonians with real spectra. We will show that such states are available in a complex SSH model, which can be experimentally realizable with current technology.

\section{Formalism}

According to the bulk-boundary correspondence, topological edge states appear at the interface of two topologically distinct Hermitian systems. For example, such states appear at open edges of a topologically non-trivial tight binding lattice, since the air is topologically trivial. Topological edge states can also appear at domain wall, which occurs if some discrete symmetries are broken along the 1D tight binding lattice.\\
Let us now study topological edge states that appear at the interface of non-Hermitian systems. Non-Hermitian Hamiltonians considered here are assumed to be $\mathcal{PT}$ symmetric and have gapped real spectra. We consider two types of interface. \\
{\it{ i-) the interface formed between topologically distinct two non-Hermitian systems}} : Since the two non-Hermitian systems are assumed to be topologically inequivalent, there is no way to adiabatically deform these two non-Hermitian Hamiltonians into each other without band gap closing. Now, a question arises. Do exceptional points occur during the topological phase transition of these two systems? If they do, the corresponding eigenvalues and the eigenstates coalesce. Generally speaking, the answer is yes but we emphasize that there may exist some systems where band gap closing points are not exceptional points. It would be interesting to find such a system. Assume that exceptional points occur during the phase transition. Then complex energy eigenvalues appear as we keep adiabatically deforming the non-Hermitian Hamiltonian. In other words, $\mathcal{PT}$ symmetry is spontaneously broken when deforming these two $\mathcal{PT}$ symmetric non-Hermitian Hamiltonians into each other. This implies that the two systems have real valued energy eigenvalues in their bulk but the topological edge states at the interface have complex energy eigenvalues. Therefore, such an interface can be used to obtain topological laser \cite{lasing}.\\
{\it{  ii-) the interface formed between topologically distinct non-Hermitian and Hermitian systems}} : A non-Hermitian system with open boundaries falls into this category since the air is a topologically trivial insulator. Since they are not topologically equivalent, there exists s no adiabatic deformation connecting the two Hamiltonians. In other words, somewhere along the way, band gap must close and reopens. Band gap closing occurs during adiabatic deformation of the non-Hermitian Hamiltonian into the Hermitian one. As opposed to the previous case, the non-Hermitian degree must be adiabatically switched-off, too.  Generally speaking, exceptional points occurs and then complex energy eigenvalues appear during the phase transition. Fortunatelly, exceptional point are less likely to occur compared to the previous case. In this way, stable topological edge states (i. e., edge states with real eigenvalues) appear at the interface. Stable topological edge states can exists in a complex extension of the SSH lattice as theoretically shown in \cite{cem0001} and experimentally realized in \cite{sondeney1}.
\begin{figure}[t]\label{2678ik0}
\includegraphics[width=9cm]{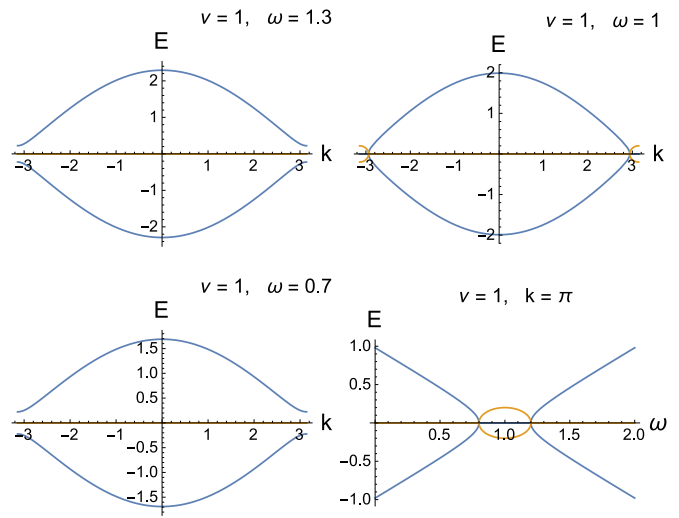}
\caption{ The energy bands for $\gamma=0.2$, where the curves in blue and orange represent real and imaginary part of the energy eigenvalues, respectively. At the top plots, the energy eigenvalues are real and they are gapped. However, the two cases are topologically distinct. Whenever phase transition occurs, i. e., the band gap closes, the system enters broken $\mathcal{PT}$ symmetric region since complex spectrum appears. Therefore, at the interface of two systems with $\ds{\omega=1.3}$ and $\ds{\omega=0.7}$, there exists complex energy interface states. The right plot down shows the interval of $\omega$ for the appearance of complex energy eigenvalues. }
\end{figure}
\subsection{Non-Hermitian SSH Model}
So far, we have discussed qualitatively the existence of complex edge states at the interface of two topologically inequivalent $\mathcal{PT}$ symmetric non-Hermitian systems with real spectra. To illustrate our idea, consider the following complex extension of the celebrated SSH model, which is a one dimensional tight-binding model with alternating hopping amplitudes and gain and loss
\begin{equation}\label{mczclaz}
H(k)=\left(\nu+\omega\cos(k) \right)~\sigma_x+\omega\sin(k)~\sigma_y+i~\gamma~ \sigma_z
\end{equation}
where $\vec{\sigma}$ are Pauli matrices, the crystal momentum $k$ runs over the first Brillouin zone, $-\pi<k<\pi$, the real-valued positive parameters $\ds{\nu>0}$, $\ds{\omega>0}$ are tunneling amplitudes and $\ds{\gamma}$ is the non-Hermitian strength. The corresponding energy eigenvalues are given by $\ds{E_{\mp}=\mp\sqrt{ \nu^2+\omega^2+2~\nu~\omega  \cos(k)-\gamma^2  }}$. Consider first the Hermitian limit, $\ds{\gamma=0}$, in which two bands are symmetrically arranged about zero energy and separated by a gap of $\ds{|\omega-\nu|}$. If we deform the Hamiltonian by varying $\ds{\omega }$ from a value $\ds{\omega>\nu}$ to a value $\ds{\omega<\nu}$ for fixed $\ds{\nu}$, we see that the band gap closes and reopens at $\ds{\omega=\nu}$. This shows us that topological phase transition occurs exactly at $\ds{\omega=\nu}$. Therefore, the cases with $\ds{\omega>\nu}$ and $\ds{\omega<\nu}$ are topologically distinct. In the non-Hermitian case, $\gamma\neq 0$, the band gap gets narrower with $\ds{\gamma}$ at fixed $\ds{\nu}$. The band gap closes and topological phase transition occurs when $\ds{\omega=\nu+\gamma}$. Contrary to the Hermitian case, exceptional points occur when the band gap is zero. More precisely, the two bands coalesce at $\ds{k=\mp\pi}$ in such a way that eigenvalues and eigenstates become simultaneously degenerate. In the Hermitian limit, band gap reopens just after closing (if $\omega$ is decreased infinitesimally). However, this is not the case in our non-Hermitian problem. If we decrease $\ds{\omega}$ below than the critical value $\nu+\gamma$, complex energy eigenvalues appear in pairs and the real part of the band gap remains zero until $\ds{\omega}$ is equal to $\nu-\gamma$. Note that at $\ds{\omega=\nu}$, the imaginary part of the energy eigenvalues takes its maximum value. At $\ds{\omega=\nu-\gamma}$, exceptional points occur once more and the band structure becomes real-valued again. The band gap reopens if we decrease $\ds{\omega}$ further. In other words, $\mathcal{PT}$ symmetry is spontaneously broken in the interval $\ds{\nu-\gamma<\omega<\nu+\gamma}$ while it is not broken and the spectrum is real valued out of this interval. In Fig.1, we plot the band structure for three choices of the parameter $\ds{\omega}$ for fixed $\ds{\gamma=0.2}$ and $\ds{\nu=1}$. One can also see the reality of the energy band as a function of $\omega$ from the figure. \\
Our above discussion was for the infinitely extended periodical complex SSH system. Let us now study topological edge states and their stabilities for a finite chain. Consider now two topologically distinct complex SSH systems. The two systems are assumed to have real-valued gapped spectra. Consider now that an interface is formed between these two topologically distinct systems. Exceptional points must be crossed somewhere along the way and we expect topological edge states with complex energy eigenvalues at the interface. We emphasize that topological edge states occur not only at the interface but also at the open end of the topologically nontrivial chain. These two topological edge states are well localized and robust against disorder. \\
To validate our discussion, we perform numerical computation for a finite chain of $N=20+20=40$ lattice sites. The chain-I and chain-II consist of $20$ lattice sites and these two chains are coupled together with a coupling constant ${\Delta}$. The Fig.2 illustrates such a system for $N=6+6=12$. To make our system non-Hermitian, we introduce balanced gain and loss into the system. One can introduce alternating gain and loss along the whole lattice. This leads to a very small critical value of $\ds{\gamma}$ for $\mathcal{PT}$ symmetry breaking. Instead we introduce gain and loss at the neighboring sites of the edges as can be seen from the Fig.2. In this way,  one can study the system in the unbroken $\mathcal{PT}$ symmetric region in a wider range of non-Hermitian strength. We note that gain and loss are not introduced at edges since they have detrimental effect on topological edge states as discussed in \cite{cem0001}. The non-Hermitian strength is assumed to be equal to $\ds{\gamma=0.5}$. We first numerically check that both chains have real spectra in the limit ${\Delta=0}$. This is because of the fact that both systems have $\mathcal{PT}$ symmetry. Suppose now that they are coupled together with a small coupling constant $\ds{\Delta=0.1}$. Switching-on $\Delta$ breaks the $\mathcal{PT}$ symmetry spontaneously at  the interface as a result of topological phase transition. Therefore we expect that the energy eigenvalue of the topological edge state at the interface becomes complex-valued. To see topological phases for various values of tunneling parameter, we parametrize the tunneling parameters as $\ds{\omega=1+0.5 \cos(\Phi)}$ and $\ds{\nu=1-0.5 \cos(\Phi)}$, where modulation phase $\ds{\Phi}$ is another degree of freedom. The Fig.3 plots the real and the imaginary parts of the spectrum as the parameter $\ds{\Phi}$ is varied. As seen, the imaginary part of the coupled system is different from zero at all values of $\Phi$. This is in agreement with our above discussion. If we look at the real part of the energy spectrum in the figure-3, we see that topological zero energy modes appears in the system unless $\ds{\Phi}$ is not inside two intervals around $\ds{\Phi=\pi/2}$ and $\ds{\Phi=3\pi/2}$ (the tunneling becomes not staggered, $\ds{\omega=\nu}$, at $\ds{\Phi=\pi/2}$ and $\ds{\Phi=3\pi/2}$). Therefore, we say that the system is topologically trivial and  localized edge states don't exist only in these two small intervals. Below, we study topological edge states. \\
\begin{figure}[t]\label{2678ik1}
\includegraphics[width=9cm]{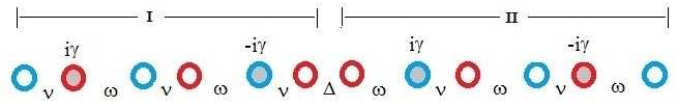}
\caption{ Our structure is displayed for $N=6+6=12$ lattice sites. The gain and loss (shaded circles) are assumed to be located as in the figure. The non-Hermitian degree is chosen in such a way that both chains have real spectra. The first 6 sites are in the chain-I region while the second 6 sites are in the chain-II region. Both chains are SSH systems so tunneling amplitudes alternates from site to site. But they are topologically distinct. The two chains are coupled with a tunneling amplitude $\ds{\Delta}$. Therefore topological phase transition occurs and topological edge states with complex energy appear at the interface. Note also that topological edge states occurs at the open end of topologically non-trivial chain, too.}
\end{figure}
\begin{figure}[t]
\includegraphics[width=9cm]{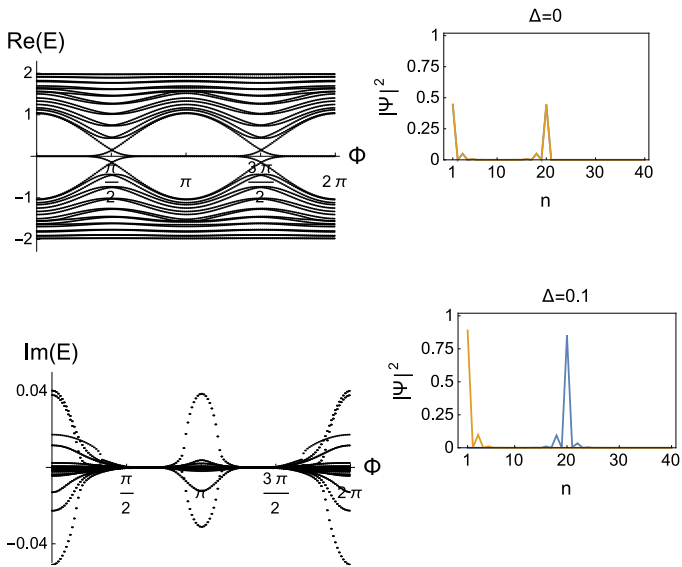}
\caption{ The real and imaginary parts of the energy eigenvalues for two topologically distinct complex SSH chains with $N=20+20=40$ and $\gamma=0.5$. The two chains are coupled together with $\ds{\Delta=0.1}$. The hopping amplitudes are parametrized as $\ds{\nu=1-0.5 \cos(\Phi)}$ and $\ds{\omega=1+0.5 \cos(\Phi)}$, where $\Phi$ is an additional degree of freedom. The coupled system has topological zero energy modes for all $\Phi$ unless $\Phi$ is not around either $\ds{\pi/2}$ or $\ds{3\pi/2}$. We also plot the densities of topological edge states for $\Delta=0$ and $\Delta=0.1$. The stable edge states are depicted in orange color while the edge state with complex energy are in blue color. As can be seen, topological edge state with complex energy eigenvalue appear at the interface when $\Delta=0.1$.}
\end{figure}
Let us numerically obtain edge states at a specific value of the modulation phase $\ds{\Phi=0}$. The chain-I is topologically nontrivial and topological edge states appear at both edges of this chain. If the two chains are not coupled, $\Delta=0$, there are two zero energy topological states and each one is symmetrically arranged on both open edges. Therefore, the density has two peaks at both ends and the peak value is around $0.4$. This can be seen from the Fig-3. We stress that these zero energy topological edge states are stable since their energy eigenvalues have no imaginary parts. Consider now that the two chains are coupled with $\ds{\Delta=0.1}$. In this case, the symmetry at both edges of the chain-I is lost since one of its edge has open boundary while the other one has an interface with the non-Hermitian topologically trivial system. In this case, topological edge states at the two edges of the chain-I are not mixed any more. Therefore, the corresponding maximum density becomes approximately equal to $0.8$ as can be seen from  the Fig-3. As expected, the edge state at the open edge is stable (it has real valued energy eigenvalue) while the edge state at the interface is either growing or decaying depending on the sign of $\gamma$. Note that the edge state at the interface can be made growing or decaying by changing the sign of $\ds{\gamma}$, which can be achieved by interchanging the gain and loss locations. As discussed above, the complex nature of the energy eigenvalue for the edge states at the interface is due to the fact that exceptional points are crossed during the topological phase transition.\\
The most interesting feature of the topological edge states is that they are robust against certain types of disorder in the system. We analyze robustness of the edge states in our system against  tunneling amplitude disorder. In our numerical computation, we introduce randomized coupling all over the lattice. The new tunneling amplitudes become $\omega\rightarrow\omega+\epsilon_n$ and $\nu\rightarrow\nu+\delta_n$, where $\epsilon_n$ and $\delta_n$ are real-valued random set of constants with $|\epsilon_n|<<\omega$ and $|\delta_n|<<\nu$. Therefore, the tunneling amplitudes between $n$-th and $n\mp1$-th sites become completely independent. In our numerical computation we take $-0.1<\epsilon_n<0.1$ and $-0.1<\delta_n<0.1$. We perform numerical calculation and repeat it for 300 different random numbers to study topological robustness against the disorder. In each calculation, we find that the real part of the energy eigenvalues of these edge states resist the disorder, i. e., they are always equal to zero. This is expected because of the topological nature of the edge states. However, the energy eigenvalues for the bulk states change considerably with disorder. The imaginary parts of the energy eigenvalues of both topological and bulk states change with $\epsilon_n$ and $\delta_n$ since the disorder breaks the $\mathcal{PT}$ symmetry of the system. In the absence of disorder, the imaginary parts of the energy eigenvalues of the edge states are equal to $20 \times10^{-4}$ and $25\times10^{-8}~$ at the open edge and the interface, respectively. Note that the maximum absolute value of the imaginary parts of the energy eigenvalues of the bulk states is $4\times10^{-2}$. We can say that the topological edge state at the interface is practically stable. This is also true in the presence of disorder. We calculate the average value and root mean square deviation of $Im(E)$ for the topological state at the open edge in the presence of the disorder. They are given by $29 \times10^{-4}$ and $24 \times10^{-4}$, respectively. Let us now study eigenstates in the presence of disorder. The eigenstates and the corresponding densities for the bulk states are highly sensitive to disorder. However, topological edge states remain well localized around the edges even in the presence of disorder. In fact, the densities of topological edge states with disorder are almost the same as the one given in the Fig-3 for $\Delta=0.1$. This shows us that topological edge states are immune to disorder.\\
One can find other examples in $1D$ or higher dimensions to explore complex topological edge states. But the above example is particularly interesting since the complex SSH Hamiltonian (1) can be realizable in photonics using waveguides. An experiment similar to the one \cite{sondeney1} can verify our findings.
\section{Conclusion}
In this paper, we have studied topological edge states in non-Hermitian systems. Although bulk-boundary correspondence is well understood in Hermitian systems, its complex extension is still absent in non-Hermitian systems. In the literature, there is no general theory explaining topological edge states at the interface of two topologically distinct non-Hermitian systems. In this paper, we have discussed this issue and given an example. We have shown that exceptional points are crossed somewhere along the way during non-Hermitian topological phase transition. This in turn leads to topological edge states with complex energy eigenvalue even if the two topologically distinct non-Hermitian systems have real-valued gapped spectra. Our system may have applications in a topological laser system. It is worth studying higher dimensional topological phase transition in non-Hermitian systems.\\
This study is supported by Anadolu University Scientific Research Projects Commission under the grant no: 11705F208

\end{document}